\def\rn{\noindent\parshape 2 0truecm 8.5truecm 0.3truecm 8.2truecm}
\def\rn{}
\def\nn#1 #2{#2. #1}				
\def\nnn#1 #2 #3{#2. #3. #1}			
\def\nnnn#1 #2 #3 #4{#2. #3. #4 #1}		
\def\nnnnn#1 #2 #3 #4 #5{#2. #3. #4 #5. #1}	
\def\dualand{ and\hbox{ }}				
\def\multiand{, and\hbox{ }}				
\def\rf#1;#2;#3;#4;#5 {{\frenchspacing\par\rn#1, #3 {\bf #4}, #5 (#2). \par}}
\def\rg#1;#2;#3;#4;#5;#6 {{\frenchspacing\par\rn#1, #3 {\bf #4}, #5 (#2). \par}}
\def\rfbook#1;#2;#3;#4;#5 {{\frenchspacing\par\rn#1, {\it #3} (#5, #4, #2).\par}}
\def\rfprep#1;#2;#3 {{\par\frenchspacing\rn#1, #3 (#2).\par}}

\def\eV{{\rm eV}}

\def\expec#1{\langle#1\rangle}

\def\etal{{\frenchspacing\it et al.}}
\def\ie{{\frenchspacing\it i.e.}}
\def\eg{{\frenchspacing\it e.g.}}
\def\etc{{\frenchspacing\it etc.}}

\def\beq#1{\begin{equation}\label{#1}}
\def\eeq{\end{equation}}
\def\beqa#1{\begin{eqnarray}\label{#1}}
\def\eeqa{\end{eqnarray}}
\def\eq#1{equation~(\ref{#1})}
\def\Eq#1{Equation~(\ref{#1})}
\def\eqn#1{~(\ref{#1})}

\def\fig#1{Figure~\ref{#1}}
\def\Fig#1{Figure~\ref{#1}}

\def\sec#1{Section~\ref{#1}}

\def\spose#1{\hbox to 0pt{#1\hss}}
\def\simlt{\mathrel{\spose{\lower 3pt\hbox{$\mathchar"218$}}
     \raise 2.0pt\hbox{$\mathchar"13C$}}}
\def\simgt{\mathrel{\spose{\lower 3pt\hbox{$\mathchar"218$}}
     \raise 2.0pt\hbox{$\mathchar"13E$}}}
\def\simpropto{\mathrel{\spose{\lower 3pt\hbox{$\mathchar"218$}}
     \raise 2.0pt\hbox{$\propto$}}}

\def\ed{\end{document}}

\def\draft{
}

\def\l{\ell}

\def\m{{\bf m}}
\def\n{{\bf n}}
\def\p{{\bf p}}

\def\x{{\bf x}}
\def\y{{\bf y}}
\def\z{{\bf z}}

\def\xt{\tilde{\x}}

\def\bzero{{\bf 0}}
\font\bfmath=cmmib10
\def\err{\hbox{\bfmath\char'042}}	

\def\I{{\bf I}}

\def\N{{\bf N}}

\def\Q{{\bf Q}}

\def\SS{{\bf S}}

\def\Sig{{\bf\Sigma}}

\def\W{{\bf W}}
\def\tr{\hbox{tr}\,}

\def\lmax{\l_{\rm max}}
\def\zz#1#2#3{$#1^{+#2}_{-#3}$}
\def\zzp#1#2{$#1^{+#2}$}

\def\Cuc{\N^{\rm (meas)}}
\def\Cic{\N^{\rm (ical)}}
\def\Csc{\N^{\rm (scal)}}
\def\Cbeam{\N^{\rm (beam)}}
\def\dT{\delta T}

\def\Ob{\Omega_{\rm b}}
\def\Oc{\Omega_{\rm cdm}}
\def\Od{\Omega_{\rm dm}}
\def\Ok{\Omega_{\rm k}}
\def\Ol{\Omega_\Lambda}
\def\Om{\Omega_{\rm m}}
\def\On{\Omega_\nu}
\def\ob{\omega_{\rm b}}

\def\od{\omega_{\rm dm}}

\def\on{\omega_\nu}
\def\fn{f_\nu}

\def\dT{\delta T}

\def\ns{n_s}
\def\nt{n_t}
\def\As{A_s}

\def\zion{z_{ion}}

\def\l{\ell}

\documentstyle[prd,aps,epsf]{revtex}
\draft
\begin{document}
\twocolumn[\hsize\textwidth\columnwidth\hsize\csname@twocolumnfalse\endcsname



\title{Is cosmology consistent?}

\author{Xiaomin Wang$^1$, Max Tegmark$^1$ \& Matias Zaldarriaga$^2$}

\address{$^1$Dept. of Physics, Univ. of Pennsylvania, Philadelphia, PA 19104;
  xiaomin@hep.upenn.edu}

\address{$^2$Dept. of Physics, New York University,  
4 Washington Pl., New York, NY 10003}

\date{May 10 2001. Submitted to Phys. Rev. D.}

\maketitle 

\begin{abstract} 
We perform a detailed analysis of the latest CMB measurements 
(including BOOMERaNG, DASI, Maxima and CBI),
both alone and jointly with other cosmological data
sets involving, e.g., galaxy clustering and the Lyman Alpha Forest.
We first address the question of whether the CMB data are internally 
consistent once calibration and beam uncertainties are taken into account,
performing a series of statistical tests. With a few minor caveats,
our answer is yes, and we compress all data into a single set of
24 bandpowers with associated covariance matrix and window functions. 

We then compute joint constraints on the 11 parameters of the 
``standard'' adiabatic inflationary cosmological model.
Out best fit model passes a series of physical consistency checks and
agrees with essentially all currently available cosmological data.
In addition to sharp constraints on the cosmic matter budget
in good agreement with those of the BOOMERaNG, DASI and Maxima teams, 
we obtain a heaviest neutrino mass range $0.04-4.2$ eV and 
the sharpest constraints to date on gravity waves which
(together with preference for a slight red-tilt) 
favors ``small-field'' inflation models.
\end{abstract}
\bigskip
] 


\section{Introduction} 

The recent discovery 
\cite{Netterfield01,Halverson01,Lee01}
of multiple peaks in the cosmic
microwave background (CMB) power spectrum 
marks a major milestone in cosmology.
Confirming 1970 predictions of 
Peebles \& Yu \cite{PeeblesYu70} and Sunyaev \& Zeldovich \cite{Sunyaev70},
it greatly bolsters the credibility of the emerging 
standard model of cosmology, and allows many of its free parameters
to be measured with a precision that cosmologists
have yet to get accustomed to.

This new precision also offers new opportunities for consistency
tests, both for systematic errors that might affect individual
measurements and for incorrect assumptions about the underlying physical
processes. The goal of the present paper is to 
carry out these two types of tests.

We begin in \sec{SystematicsSec} by testing measurements of the CMB 
power spectrum for consistency with a 
series of quantitative statistical tests, including 
the effects of calibration and beam uncertainties.
Since the customary plot of available measurements 
has now evolved into a bewildering zoo of over 100 band power measurements 
which is increasingly difficult to interpret visually
because of calibration and beam effects, we
perform an essentially lossless data compression of all data
into a single set of 24 bandpowers with associated covariance 
matrix and window functions, useful as a starting point for parameter fitting. 

In \sec{ParameterSec}, we compute quantitative constraints on 
the 11 parameters of the 
``standard'' adiabatic inflationary cosmological model in a variety of 
different ways, using data from, \eg,  the CMB, galaxy clustering,
the Lyman Alpha Forest, Big Bang nucleosynthesis (BBN) and 
Hubble constant measurements in various combinations.
This enables us to identify a number of parameter constraints 
that are robust and consistent with all data, as well as areas
where there is slight tension between data sets pulling in
different directions.
Although numerous such studies have been performed in the recent
literature, \eg, 
\cite{Lange00,boompa,Bambi00,observables,Jaffe00,Padmanabhan00,Lineweaver00,Kinney01,Hannestad01,Hannestad01b,Griffiths01,Phillips01,BondConfProc,CosmicTriangle,Novosyadlyj00,Novosyadlyj00b,Durrer00,Bridle00,Turner01,Holder01},
the dramatically improved precision allowed by 
new BOOMERaNG \cite{Netterfield01}, DASI \cite{Halverson01}, 
Maxima \cite{Lee01} and CBI \cite{Padin01} data makes
it worthwhile and timely to revisit this issue\footnote{Numerous additional
multiparameter studies were submitted after the present paper,
the most similar in focus being those by the 2dF redshift survey team
\cite{Efstathiou01,Lahav01}.
}.
The present work extends the solid recent analyses of the experimental 
teams \cite{Netterfield01,Pryke01,Stompor01} mainly in the following ways:
\begin{enumerate}
\item Inclusion of more parameters allows us to place constraints on
neutrinos and gravity waves and to quantify the additional degeneracies that
they introduce.
\item Joint analysis of all CMB data allows us to place stronger constraints 
and perform consistency tests.
\item Inclusion of explicit power spectrum modeling for the 
galaxy clustering and for the Lyman Alpha Forest allows stronger constraints
and important new cross-checks.
\end{enumerate}

\section{Is the CMB story consistent? Comparing and combining measurements of the
angular power spectrum}
\label{SystematicsSec}

In this section we test the CMB data for internal 
consistency and combine them into a single set of band powers,
calibrating the experiments against each other. 

\subsection{CMB data}

\begin{figure}[tb] 
\vskip-1.2cm
\centerline{\epsfxsize=9.0cm\epsffile{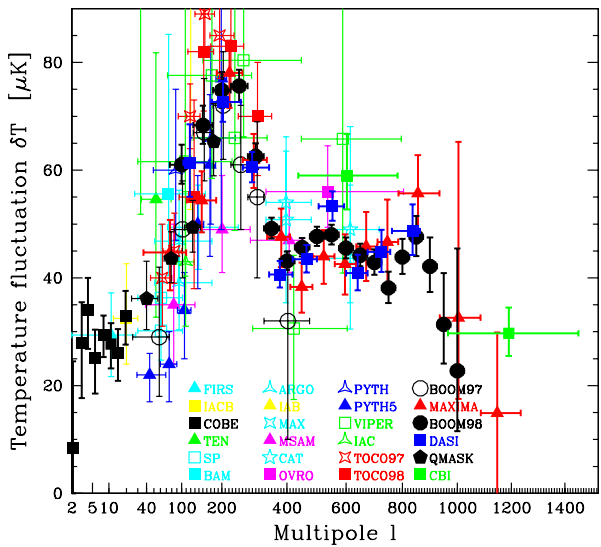}}
\vskip-0.8cm
\smallskip
\caption{\label{cmbdataFig}\footnotesize%
CMB data used in our analysis.
Error bars do not include calibration or beam errors
which allow substantial vertical shifting and 
tilting for some experiments.
}
\end{figure}

\begin{figure}[tb] 
\vskip-1.2cm
\centerline{\epsfxsize=9.0cm\epsffile{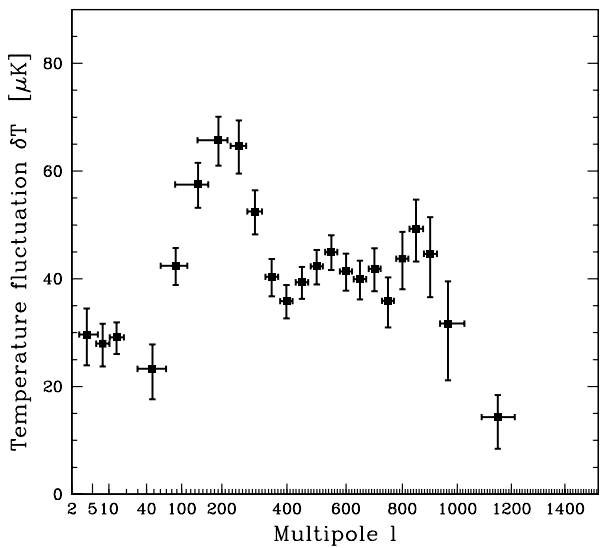}}
\vskip-0.8cm
\smallskip
\caption{\label{binnedFig}\footnotesize%
Combination of all data from \protect\fig{cmbdataFig}.
These error bars include the effects of beam and
calibration uncertainties, which cause long-range correlations
of order 5\%-10\% over the peaks. In addition, points tend to be anti-correlated 
with their nearest neighbors, typically at the level of 
a few percent.
The horizontal bars give the characteristic widths of the window functions (see text).
}
\end{figure}

\Fig{cmbdataFig} shows the 105 band power measurements used in our
analysis.
Starting with the data tabulated in \cite{10par}, 
we have added the new measurements from
CBI \cite{Padin01}, 
QMASK \cite{qmask}, 
BOOMERaNG \cite{Netterfield01}, 
DASI \cite{Halverson01} and 
Maxima \cite{Lee01}.
Since QMASK combines the Saskatoon \cite{Netterfield97} and 
QMAP \cite{qmap1,qmap2,qmap3} datasets,
these have been omitted. A recent data review is given in \cite{Gawiser00}.

The success of experimental CMB work has made data plots
such as \fig{cmbdataFig} increasingly bewildering 
and difficult to interpret. Not only do many 
data points with widely different error bars overlap,
but important correlations due to calibration and beam uncertainties
are difficult to visualize graphically and tend not to be included
in the plotted error bars.
The obvious solution to this problem is some sort of data compression.

A radical but common example of this is to simply throw away most 
of the data and show/analyze only one or two experiments, often the most
recent.
This is not ideal, however, since it both wasteful of information
and lacks an objective criterion for data culling. 
Moreover, among the most accurate and thoroughly 
systematics-tested measurements on large scales still come from older maps
(COBE DMR \cite{Bennett96} and QMASK \cite{qmaskpow}).

A more desirable alternative is to average the data together somehow,
into a single measurement of power on each angular scale.
Such CMB data compression has been performed by many authors, \eg,
\cite{Jaffe00,10par,ScottWhiteSilk95,Pierpaoli00,KnoxPage00},
and can retain all cosmological information provided that the new
power bins are narrow enough to not smooth out important power spectrum
features. However, such compression throws away any evidence for
discrepancies between experiments that may have been present in the
full data set, so it is important to complement the averaging by
consistency checks.

\subsection{Combining experiments}

Let us group the power measurements at hand (say the 105 measurements
of $\delta T^2$ from \fig{cmbdataFig}) into a vector $\y$.
We will model this as
\beq{xModelEq}
\y=\W\x+\n,
\eeq
where $\x$ is a vector containing the true power spectrum coefficients 
$\delta T_\l^2$ up to some sufficiently large multipole $\lmax$,
the window function matrix $\W$ encodes the angular sensitivity 
of the measurements (the rows of $\W$ sum to unity) and the noise vector 
$\n$ represents all forms of measurement error. 
We model the errors as random with zero mean
($\expec{\n}=\bzero$) and with a covariance matrix 
$\N\equiv\expec{\n\n^t}$ that is the sum of four terms, 
\beq{Neq}
\N=\Cuc + \Csc + \Cic + \Cbeam, 
\eeq
corresponding to basic measurement errors, 
source calibration errors, 
instrument calibration errors
and beam errors, respectively.
In general, all of these errors depend on the actual power spectrum $\x$,
either through sample variance \cite{BJK} or because 
calibration and beam errors are multiplicative rather than additive.
Below we will make the approximation that the relative errors
are small ($\ll 1$). In this limit, $\N$ reduces to a known matrix independent 
of $\x$. Explicit expressions for the four matrices in \eq{Neq} are
given in Appendix A.

Given $\W$, $\N$ and $\y$, it is straightforward to invert 
\eq{xModelEq} to compute an estimate of the underlying power
spectrum $\x$. This problem is mathematically identical to that involved
in CMB mapmaking \cite{mapmaking,Wright96} except that the
matrices involved are small enough to be trivial to manipulate 
numerically. As our estimator of $\x$ we use
\beq{inversionEq}
\xt\equiv\left[\W^t\N^{-1}\W\right]^{-1}\W^t\N^{-1}\y,
\eeq
which can be shown to be unbiased $(\expec{\xt}=\x)$, to minimize
the rms noise in each power band and, 
if the noise properties are Gaussian, to retain all 
information about the true power spectrum $\x$ from 
the original data $\y$ \cite{mapmaking}.
The corresponding covariance matrix of the noise $\bf\epsilon\equiv\xt-\x$ is
\beq{SigcovEq}
\Sig\equiv\expec{\err\err^t} = 
\left[\W^t\N^{-1}\W\right]^{-1}.
\eeq
The resulting power spectrum $\xt$ is shown in \fig{binnedFig} and
listed in Table 1. The corresponding covariance matrix $\Sig$
is available at\\ $www.hep.upenn.edu/\sim max/cmb/experiments.html$.

When computing this spectrum, we did not treat the power as an independent
parameter at each multipole. Rather, we treated the power spectrum as piecewise 
constant, parametrized by its height $x_i$ in the 24 bands listed in Table 1.
Since our compressed band powers $\xt$ are simply linear combinations 
of the original measurements $\y$, we are able to compute their window functions
exactly by taking the same linear combinations of the rows of $\W$ from
\eq{xModelEq}. Our compressed data set can therefore be
analyzed ignoring the details of how it was constructed
(ignoring the first column of Table 1),
viewed as simply the window matrix times the true power.

The characteristic widths of these window functions are reflected by the 
horizontal 
bars in \fig{binnedFig} and listed in the Table; the exact windows
are available on the above-mentioned web 
site\footnote{
The horizontal location of a data point in \fig{binnedFig} corresponds to the
median (50\% quantile) of the absolute value of its window function.
We use absolute values to be pedantic, since some windows go slightly negative 
in places, although this makes a negligible difference for the plot.
The horizontal bars to the left and right extend to the 20\% and 80\% quantiles,
respectively, indicating the characteristic window function width.
These quantiles correspond to the full-width half-max (FWHM) for a 
Gaussian window.
}.
This correlation matrix includes the residual effects of 
calibration uncertainty and beam errors. These long-range
correlations are found to be moderate, 
typically of order 5-10\% over the peaks,
which shows that the experiments have to a certain extent been 
calibrated off of each other. In addition, there tends to be
a slight anti-correlation between neighboring points, typically
at the level of a few percent, as the power spectrum inversion
performs a slight deconvolution of the input window functions 
from the experiments used.

\bigskip
\bigskip
\bigskip

\bigskip
\noindent
{\footnotesize
{\bf Table 1} -- Band powers combining the
information from all 105 CMB data points
from \protect\fig{cmbdataFig}.
The 2nd column gives the medians and characteristic widths of 
the window functions as detailed in the text. The spectrum was computed treating 
$\delta T^2$ as constant in the bands listed in the 
first column.
The error bars in the 3rd column
include the effects of calibration and beam uncertainty.
The full $24\times 24$ correlation matrix and $24\times 2000$
window matrix are available at $www.hep.upenn.edu/\sim max/cmb/experiments.html$.
\bigskip
\begin{center}
{\footnotesize
\begin{tabular}{|c|c|c|}
\hline
$\l$-Band	&$\l$-window	&$\delta T^2\>[\mu $K$^2]$\\
\hline		
$2-2$		&$    2_{- 0}^{+ 0}$&$   50\pm 310$\\
$3-5$		&$    4_{- 1}^{+ 2}$&$  880\pm 308$\\
$6-10$		&$    8_{- 2}^{+ 3}$&$  782\pm 219$\\
$11-30$ 	&$   14_{- 3}^{+ 4}$&$  849\pm 171$\\
$31-75$ 	&$   48_{-18}^{+22}$&$  542\pm 231$\\
$76-125$	&$   88_{-28}^{+24}$&$ 1800\pm 292$\\
$126-175$	&$  138_{-51}^{+26}$&$ 3307\pm 480$\\
$176-225$	&$  191_{-55}^{+25}$&$ 4318\pm 597$\\
$226-275$	&$  249_{-24}^{+22}$&$ 4182\pm 637$\\
$276-325$	&$  298_{-24}^{+22}$&$ 2756\pm 428$\\
$326-375$	&$  351_{-21}^{+20}$&$ 1630\pm 279$\\
$376-425$	&$  398_{-21}^{+20}$&$ 1286\pm 221$\\
$426-475$	&$  450_{-21}^{+21}$&$ 1549\pm 232$\\
$476-525$	&$  499_{-21}^{+21}$&$ 1789\pm 270$\\
$526-575$	&$  549_{-21}^{+21}$&$ 2021\pm 290$\\
$576-625$	&$  600_{-21}^{+21}$&$ 1711\pm 284$\\
$626-675$	&$  649_{-22}^{+21}$&$ 1594\pm 285$\\
$676-725$	&$  701_{-21}^{+22}$&$ 1752\pm 332$\\
$726-775$	&$  749_{-21}^{+21}$&$ 1290\pm 330$\\
$776-825$	&$  800_{-22}^{+22}$&$ 1912\pm 462$\\
$826-875$	&$  850_{-24}^{+25}$&$ 2428\pm 563$\\
$876-925$	&$  900_{-23}^{+24}$&$ 1993\pm 653$\\
$926-1025$	&$  966_{-29}^{+60}$&$ 1004\pm 557$\\
$1026-\infty$	&$ 1149_{-60}^{+63}$&$  205\pm 134$\\
\hline		
\end{tabular}
}
\end{center}
}


One interesting feature of \fig{binnedFig} is that it shows
both the first and second peak\footnote{One of the 
Argentinian authors feels that there is still no conclusive evidence 
for multiple peaks in the power spectrum, especially 
in regard to any bets or wagers that may or may not be 
outstanding.}
somewhat lower than a large fraction of the
data. Indeed, the preferred recalibrations for all five 
multiband experiments flagged in the following subsection are
downward.
To understand the origin of this effect, we performed a series
of tests where the combined spectrum was recomputed with one or more
experiments omitted. The explanation centers around the BOOM98 data, 
which combines a sharp constraint on the relative heights of the first 
two peaks (even taking
the beam uncertainty into account) with a relatively large overall 
calibration uncertainty. Since both Maxima and DASI (with one exception) 
have points with small error bars below BOOM98 around the second peak, they
pull the BOOM98 calibration down. QMASK also pulls
BOOM98 down because of its statistical weight around $\l\sim 100$.
Although \fig{cmbdataFig} suggests that influence from, \eg, TOCO around
the first peak might raise BOOM98, this pull is weaker because of 
TOCO calibration uncertainties
and since error bars are overall smaller out at the second peak. 
This somewhat low normalization persists even if any one of 
Maxima, DASI or QMASK is excluded from the analysis.

\subsection{Comparing experiments}

Since the combined power spectrum presented above is only as reliable as the data
that went into it, let us now test this data for internal consistency.

As mentioned above, \eq{xModelEq} is analogous to the CMB mapmaking problem,
which means that all methods developed for comparing and combining 
maps can be applied to comparing and combining power spectra as well.
Given two power spectrum measurements $\y_1$ and $\y_2$ modeled as in \eq{xModelEq},
\eg, as
\beq{xModelEq2}
\y_1=\W_1\x+\n_1,\quad \y_2=\W_2\x+\n_2,
\eeq
we wish to know whether they are consistent
or display evidence of systematic errors. Specifically, is there 
some underlying power spectrum $\x$ such that the data sets $\y_1$ and $\y_2$ 
are consistent with \eq{xModelEq2}?
Let us consider the simple case where the two experiments
have identical window functions, that is, 
$\W_1=\W_2$. 
The general case can be reduced to this one:
In practice, we start by reducing all experiments 
to the simple form $\W=\I$ using the deconvolution method described
in Appendix D of \cite{qmask}.
 
Consider two hypotheses: 
\begin{itemize}
\item[$H_0$:]The null hypothesis $H_0$ that there are no systematic errors,
so that the difference spectrum $\z\equiv\y_1-\y_2$ consists of pure noise
with zero mean and covariance matrix $\expec{\z\z^t}=\N\equiv\N_1+\N_2$.
\item[$H_1$:]The alternative hypothesis that the difference spectrum
$\z$ has the same covariance matrix $\N$ but 
a non-zero mean $\m$.
\end{itemize}
A straightforward variation of the derivation in \cite{comparing} 
shows that 
the  ``null-buster'' statistic \cite{comparing}
\beqa{NullbusterEq}
\nu &\equiv& \frac{\z^t\N^{-1}\Q\N^{-1}\z - \tr\{\N^{-1}\Q\}}
             {\left[2\>\tr\{\N^{-1}\Q\N^{-1}\Q\}\right]^{1/2}},\nonumber\\
\Q  &\equiv&\m\m^t,
\eeqa
rules out the null hypothesis $H_0$ with the largest average
significance $\expec{\nu}$ if $H_1$ is true, and can be interpreted as
the number of ``sigmas'' at which $H_0$ is ruled out \cite{comparing}.
The case derived in \cite{comparing} differed in that the mean vanished 
under $H_1$ but that the covariance matrix contained extra signal $\SS$
--- the result was of the same form as above, but with $\Q=\SS$.
Note that for the special case $\Q\propto\N$, it reduces to simply
$\nu=(\chi^2-n)/\sqrt{2n}$, where 
$\chi^2\equiv\z^t\N^{-1}\z$ is the standard chi-squared statistic.
The null-buster test can therefore be viewed as a generalized 
$\chi^2$-test which places more weight on those particular 
modes where the expected signal-to-noise is high.
It has proven successful comparing both 
microwave background maps \cite{qmask,qmap1,qmap2,qmap3,qmaskpow} and  
galaxy distributions \cite{r,EfstathiouNullbuster}.
Tips for rapid implementation in practice are given in \cite{qmask}.

\Eq{NullbusterEq} shows that in our case, all weight is placed on a single mode $\m$. 
More generally, the test pays the greatest attention to those eigenvectors of
$\Q$ whose eigenvalues are large.
Consider first the case of calibration errors. 
Then the two measured power spectra are generically 
expected to have the same shape but different normalizations, 
so that the vectors $\expec{\y_1}$ and $\expec{\y_2}$ 
are parallel but with different lengths.
In other words, $\SS=\bzero$ and $\m=\expec{\z}\propto\W\x$, 
so the mode $\m$ that we want our test to be sensitive to is shaped like the
the expected sky power spectrum itself --- we make this choice below.
Similarly, beam errors show up in a different mode, which
(as discussed in Appendix A) is shaped like 
the sky power spectrum times $\l^2$ to first order.

\begin{figure}[tb] 
\centerline{\epsfxsize=8.5cm\epsffile{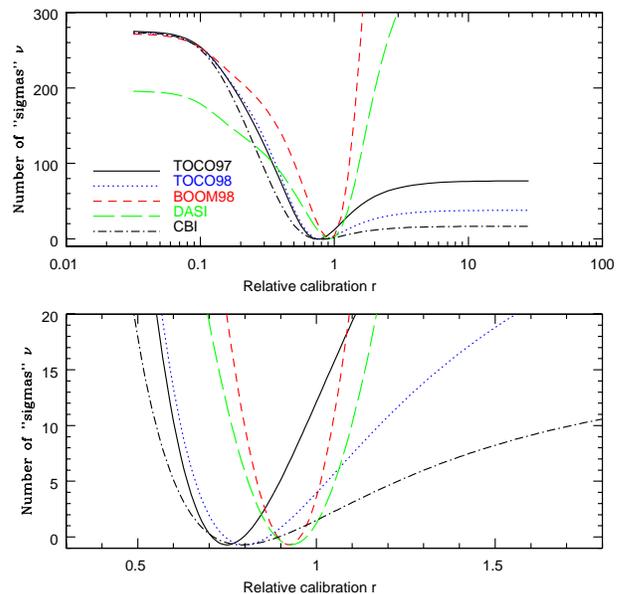}}
\smallskip
\caption{\label{nuFig}\footnotesize%
Each curve shows the number of standard deviations 
(``sigmas'') at which a given experiment is 
inconsistent with all others 
when its power spectrum $\delta T$ is multiplied by
a constant $r$. 
}
\end{figure}

\subsection{Test results}

As emphasized by, \eg, Press \cite{Press96}, it is important to
check any data collection for statistically significant outliers.  
We do this separately for each of the 24 experiments
from \fig{cmbdataFig} as follows.
We form the difference spectrum 
\beq{zEq}
\z\equiv\y_2 - r \y_1,
\eeq
where $\y_1$ and $\y_2$ are the deconvolved 
power spectra produced from the experiment under study
and from all other experiments, respectively,
and vary the normalization parameter $r$.
We take calibration and beam errors into account in computing 
the ``all other'' spectrum, but not for the experiment under consideration.
The resulting significance level $\nu$ at which the 
difference spectrum is inconsistent with zero is
plotted as a function of $r$ in \fig{nuFig}.

All experiments taken together have now detected CMB fluctuations 
at about the $270\sigma$ level, and the fact that all curves except the DASI one
asymptote to just under that value as $r\to 0$ shows that no
single experiment dominates all others in statistical weight.
To the right, as $r\to\infty$, each curve asymptotes to the 
significance level at which the experiment in question
detects signal. If the experiment under study has 
no calibration or beam errors and everything is consistent,
$\nu$ should be near zero for $r=1$, where it has vanishing mean
and unit variance ($\expec{\nu}=0$, $\Delta\nu=1$).
Only five of the 24 experiments show a significant 
difference at the $2\sigma$-level $(\nu(1)>2)$.
Their $\nu(r)$-curves are plotted in \fig{nuFig},
are seen to be perfectly consistent as well --- the
relative calibration $r$ simply has to be shifted to 
a different value, which is in all cases lower,
for which $\nu$ drops below 2.

Above we saw that the experiments could be effectively 
recalibrated off of each other.
We also computed $\nu$ for the latest BOOMERaNG data on a 
2-dimensional grid, varying both the calibration and the beam size.
However, beam information from inter-experiment
comparison is substantially less useful at the present
time than extracting the corresponding calibration 
information --- we found the BOOMERaNG 
beam constraints obtained in this way to be weaker 
than those measured from the instrument directly.

In conclusion, the quantitative tests described above show no
evidence of inconsistency between the 24 CMB experiments
when beam and calibration uncertainties are taken into account,
and the power spectrum shown in \fig{binnedFig} is consistent 
with all of them. The dip around $\l=50$, caused mainly by Python V, may warrant further
investigation to study if a smoother underlying spectrum
can be consistent with all the data in that range.

\section{Is the cosmology story consistent?
Comparing and combining different cosmological datasets}
\label{ParameterSec}

In this section we confront the CMB data with other cosmological observations, 
with the goal being both accurate constraints on cosmological parameters 
and various cross-checks on the underlying physical assumptions.
We first map out the subset of 
the 11-dimensional cosmological parameter space from \cite{concordance}
that is consistent with CMB, large scale structure (LSS) and Lyman 
Alpha Forest (Ly$\alpha$F) power spectra, with Big Bang nucleosynthesis
(BBN) and with direct Hubble constant determinations,
included and excluded in various combinations. We pay particular
attention to whether these priors are mutually consistent or
pull in different directions, both in terms of which parts of
parameter space they pull towards and in terms of how much they increase
the overall $\chi^2$.
We will return to the consistency issue 
in \sec{DiscussionSec}, comparing our ``concordance model'' 
with various other cosmological constraints.

\subsection{Analysis method}

We employ the multiparameter analysis method described in
\cite{concordance} and \cite{10par} with the following 
modifications: optional inclusion of Ly$\alpha$ power
spectra, optional discretization of the gravity wave contribution
to allow explicit limits on this parameter, inclusion of
CMB beam uncertainties as described in Appendix A and 
refined parameter grids to reflect the improved data accuracy.
It consists of the following steps:
\begin{enumerate}
\item Compute CMB, LSS and Ly$\alpha$F power spectra for a grid
of models in our 11-dimensional parameter space.
\item Compute a likelihood for each model that quantifies how well it fits the
data.
\item Perform 11-dimensional interpolation and marginalization to obtain
constraints on individual parameters and parameter pairs.
\end{enumerate}
Our 11 parameters are 
\beq{pEq}
\p\equiv(\tau,\Ok,\Ol,\od,\ob,\fn,\ns,\nt,\As,r,b).
\eeq
These are the reionization optical depth $\tau$, 
the primordial amplitudes $\As$, $r\As$ and tilts $\ns$, $\nt$ 
of scalar and tensor fluctuations, 
a bias parameter $b$ defined as the ratio between rms 
galaxy fluctuations and rms matter fluctuations on 
large scales,
and five parameters specifying the cosmic matter budget.
The various contributions $\Omega_i$ to critical density are for
curvature $\Ok$, vacuum energy $\Ol$, cold dark matter $\Oc$, 
hot dark matter (neutrinos) $\On$ and baryons $\Ob$.
The quantities
$\ob\equiv h^2\Ob$ and
$\od\equiv h^2\Od$ correspond to 
the physical densities of baryons
and total (cold + hot) dark matter 
($\Od\equiv\Oc+\On$), and $\fn\equiv\On/\Od$ is the fraction
of the dark matter that is hot.
We assume that the galaxy bias $b$ is constant on large scales 
\cite{Taruya01}
but make no assumptions about its value, 
and therefore marginalize (minimize) over this parameter
before quoting constraints on the other ten.
In the adaptive mesh spirit, we iteratively refined our parameter
grid to adequately resolve the peak of the likelihood function.
Our final parameter grids were as follows:
\begin{itemize}
\item $\tau=0, 0.05, 0.1, 0.2, 0.3, 0.5, 0.8$ 
\item $\Ol=0, 0.1, ...., 1.0$ 
\item $\Ok=\pm 1.0,\pm 0.8,\pm 0.6,\pm 0.5,\pm 0.4,\pm 0.3,\pm 0.2,$\\
      $\pm 0.1, \pm 0.03, 0$,
       truncated so that\\ $\Om\equiv 1-\Ok-\Ol\in[0.05,1]$
\item $\od=.02, .05, .08, .11, .13, .16, .20, .50$ 
\item $\ob=.003, .015, .018, .020, .022, .025, .03, .04, .07$ 
\item $\fn=0, 0.05, 0.1, 0.2, 0.3, 0.4, 0.6, 0.8, 1.0$ 
\item $\ns=0.5, 0.7, 0.8, 0.9, 1.0, 1.1, 1.2, 1.4, 1.7$ 
\item $\nt=-1.0, -0.7, -0.4, -0.2, -0.1, 0$ 
\item $r=0, 0.1, 0.2, 0.3, 0.4, 0.6, 0.8, 1.0, 1.4, 1.8, 2.5$
\item $\As$ is not discretized
\item $b$ is not discretized
\end{itemize}
The parameter $r$ was only discretized when computing constraints
involving $r$ --- it was treated as continuous 
and marginalized over analytically as in \cite{concordance}
when constraining other parameters.
We used the original CMB data set rather than the compressed one 
to compute the CMB likelihood, since step 2 above is the fastest 
of the three in our analysis pipeline anyway.

\subsection{Non-CMB data used}

As our LSS data, we follow \cite{concordance} in using the power
spectrum measured from by the IRAS PSCz survey \cite{Saunders00}
by \cite{pscz}, discarding all measurements on scales smaller than
20$h^{-1}$Mpc  ($k>0.3h$/Mpc) to be conservative.

As our Ly$\alpha$F data, we use the 13 recent power spectrum measurements
of \cite{Croft00}, with an additional 27\% calibration uncertainty
common to all points (see also \cite{McDonald00}). We compute theoretical predictions for
these 13 numbers corresponding to each of the $\sim$ 300 million models
in our database by first computing the matter power spectrum $P(k)$
as in \cite{concordance}, then shifting it vertically and horizontally
(on a log-log scale) to account for the fluctuation growth and 
Hubble parameter shift between $z\sim 2.72$ and the present epoch.
We used the grow$\lambda$ package \cite{Hamilton00} for computing 
the growth factors.

We quantitatively explore how our results are affected by adding various 
other constraints (``priors''). 
For BBN, we test the prior $\ob=0.02$ for simplicity, since the error bars
on the most recent BBN estimates $\ob=0.02\pm0.002$ (95\%)
from \cite{Burles01} are smaller than our grid spacing.
For the Hubble parameter, we test 
the prior $h=0.72\pm 0.08$ from the HST Hubble Key Project \cite{Freedman00},
assuming a Gaussian distribution for simplicity.
We also try the priors $\tau=0$, $\Ok=0$, $\fn=0$,
$r=0$ and $\ns=1$ in various combinations.

\subsection{Basic results}

\begin{figure}[tb] 
\centerline{\epsfxsize=8.5cm\epsffile{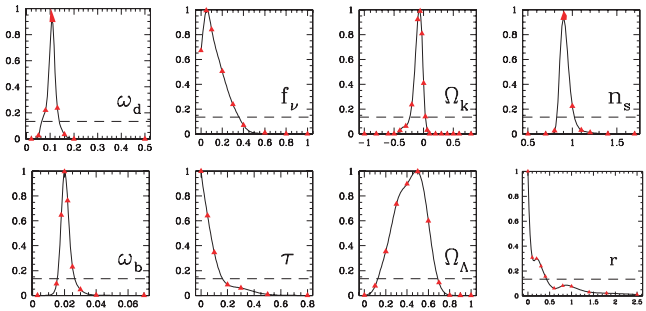}}
\bigskip
\caption{\label{1DnoFig}\footnotesize%
Constraints on individual parameters using only CMB and LSS information.
The quoted 95\% confidence limits are where each curve drops below
the dashed line.
}
\end{figure}

\begin{figure}[tb] 
\centerline{\epsfxsize=8.5cm\epsffile{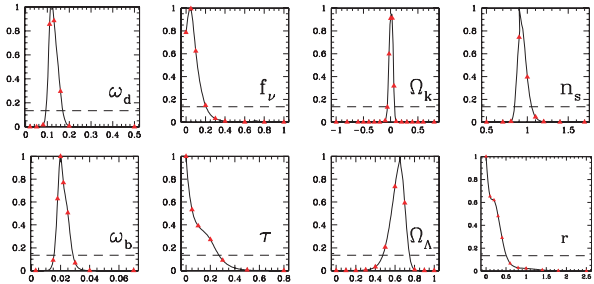}}
\vskip-3.0cm
\bigskip
\caption{\label{1DhubbleFig}\footnotesize%
Like the previous figure, but adding the 
prior $h=0.72\pm 0.08$. 
}
\end{figure}

Our constraints on individual cosmological parameters are 
listed in Table 2 for four cases.
Constraints are plotted in 
figures~\ref{1DnoFig} and~\ref{1DhubbleFig} for cases 2 and 3.
All tabulated and plotted bounds are 95\% confidence limits\footnote{
Bayesean 95\% confidence limits
are in general those that enclose 95\% of the area.
In this paper, we make the approximation that the boundary of the
confidence region is that where the likelihood has fallen by a factor
$e^{-\Delta\chi^2/2}$ from its maximum, where
$\Delta\chi^2=4$ for 1-dimensional cases (such as 
the numbers in Table 2) and $\Delta\chi^2=6.18$ for  
2-dimensional cases 
(such as figures~\ref{1DnoFig} and~\ref{1DhubbleFig}).
As shown in Appendix A of \cite{10par}, this approximation
becomes exact only for the case when the 
likelihood has a multivariate Gaussian form. 
We make this approximation to be consistent with 
the multidimensional marginalization algorithm employed
here (and by most other authors), which is equivalent to the 
integration technique only for the Gaussian case. 
}.
The first case uses constraints from CMB alone, which are 
still rather weak because of a one-dimensional degeneracy 
coupling curvature, baryons, tilt, tensors, dark matter and dark energy 
as described in the following subsection.
The second case breaks this degeneracy
by combining the CMB information with 
the power spectrum measurements from PSCz, and is seen to
give rather interesting constraints on most parameters. 
The third case adds the prior assumptions
that the Hubble parameter is $h=0.72\pm 0.08$ 
\cite{Freedman00}, which tightens up many constraints,
in particular that on $\Ol$.
The fourth case adds the assumption
that the neutrino contribution is cosmologically negligible
($\fn\sim 0$). This is equivalent to assuming that 
there is no strong mass-degeneracy between
the relevant neutrino families, and that the Super-Kamiokande atmospheric
neutrino data therefore set the scale of the 
neutrino density to be $\on\simlt 10^{-3}$ \cite{Fukuda99,Valle01}.

\bigskip
\bigskip
\noindent
{\footnotesize
{\bf Table 2} -- Best fit values and 95\% confidence limits on
cosmological parameters.
$\zion$ is the redshift of reionization and $t_0$ is the present age
of the Universe.
For the numbers above the horizontal line, 
the central values are the ones maximizing the likelihood (the best fit
model).
For the numbers below the horizontal line, 
the limits were computed from
moments as described in the text, so 
the central values are
means rather than those for the best fit model.
For instance, the Hubble parameters for the best fit models
are $h=$0.51, 0.48, 0.64 and 0.64 for the four columns, respectively, 
which differs from the mean values in the table.
\def\fnp{{$\sim$0}}
\bigskip
\begin{center}
{
\begin{tabular}{|l|c|c|c|c|}
\hline
		&CMB alone		&$+$ PSCz	&+ $h=.72\pm.08$	&+ $\fn\sim 0$\\
\hline
$\tau$		&\zzp{0.00}{.18}	&\zzp{0.00}{.16}	  &\zzp{0.00}{.27}      &\zzp{0.00}{.27}\\
$\Ok$		&\zz{-0.05}{.10}{.34}   &\zz{-0.07}{.16}{.10}     &\zz{0.02}{.05}{.08}  &\zz{0.00}{.06}{.06}\\
$\Ol$		&\zz{0.57}{.32}{.45}	&\zz{0.49}{.19}{.37}	  &\zz{0.65}{.10}{.18}  &\zz{0.66}{.09}{.14}\\
$h^2\Od$	&\zz{0.10}{.03}{.05}	&\zz{0.10}{.03}{.04}	  &\zz{0.12}{.05}{.03}  &\zz{0.12}{.04}{.03}\\
$h^2\Ob$	&\zz{0.021}{.009}{.005}	&\zz{0.020}{.006}{.005}	  &\zz{0.020}{.009}{.004}  &\zz{0.020}{.008}{.005}\\
$\fn$		&\zzp{0.08}{.59}	&\zzp{0.06}{.30}	  &\zzp{0.04}{.16}      &{\it 0}\\
$\ns$		&\zz{0.91}{.16}{.09}	&\zz{0.91}{.11}{.09}	  &\zz{0.93}{.12}{.09} &\zz{0.91}{.15}{.07}\\
\hline  	
$h$		&\zz{0.42}{.23}{.24}	 &\zz{0.57}{.31}{.30}	  &\zz{0.71}{.12}{.12}	&\zz{0.73}{.11}{.10}\\
$\zion$        	&\zz{5.7}{12.9}{5.7}     &\zz{7.2}{13.8}{7.2}     &\zz{7.6}{14.2}{7.6}  &\zz{6.8}{13.1}{6.8}\\
$t_0$ [Gyr]     &\zz{20.5}{9.0}{9.0}	 &\zz{14.2}{4.3}{4.3}  	  &\zz{12.3}{1.6}{1.6}  &\zz{12.7}{1.5}{1.5}\\
\hline		
\end{tabular}
}
\end{center}
}

For the first 7 parameters listed in Table 2, the 
numbers were computed from the corresponding
1-dimensional likelihood functions (plotted 
in \fig{1DnoFig} and \fig{1DhubbleFig} for the 
second and third cases). The best fit value corresponds to the 
peak in the likelihood function and the $95\%$ limits correspond
to where the likelihood function drops below the dashed line 
at $e^{-2}$ of the peak value.
For the remaining parameters listed,
which are not fundamental parameters in our 11-dimensional grid, 
the numbers were computed as in 
\cite{Jaffe00} 
by calculating the likelihood-weighted means and standard 
deviations over the 
multidimensional parameter space.
Here the tabulated limits are the mean $\pm 2\sigma$.

\bigskip

\begin{figure}[tb] 
\centerline{\epsfxsize=8.5cm\epsffile{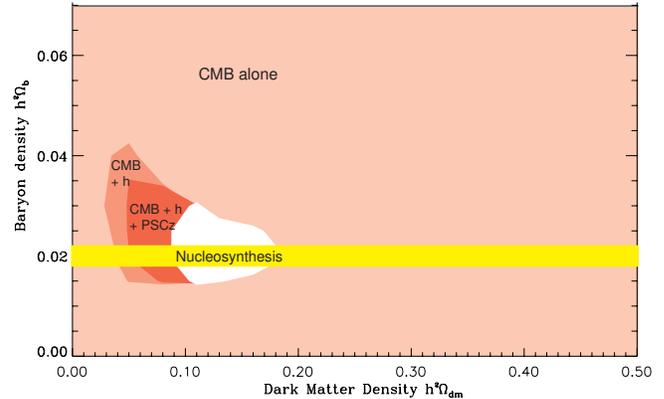}}
\bigskip
\caption{\label{odobFig}\footnotesize%
Constraints in the $(\od,\ob)$-plane. 
The large pink/grey region is ruled out by CMB data alone at
95\% confidence. The two smaller shaded regions become excluded
when imposing additional constraints from Hubble 
constant measurements ($h=0.72\pm 0.08$) and PSCz galaxy
clustering.
The yellow/light grey band shows the BBN constraints from \protect\cite{Burles01}.
The model best fitting the CMB+PSCz+$h$ constraints
has $\chi^2\approx 126$ for $127$ degrees of freedom.
}
\end{figure}

\subsection{Matter budget}

We will now investigate the parameter constraints in more detail, 
exploring which conclusions come from what assumptions. 
This subsection is
centered around the cosmic matter budget 
(the densities of baryons, cold dark matter, hot dark matter, dark energy 
and curvature) --- we return to the inflationary parameters
$\ns$, $\nt$ and $r$ in the next subsection.

\begin{figure}[tb] 
\centerline{\epsfxsize=8.5cm\epsffile{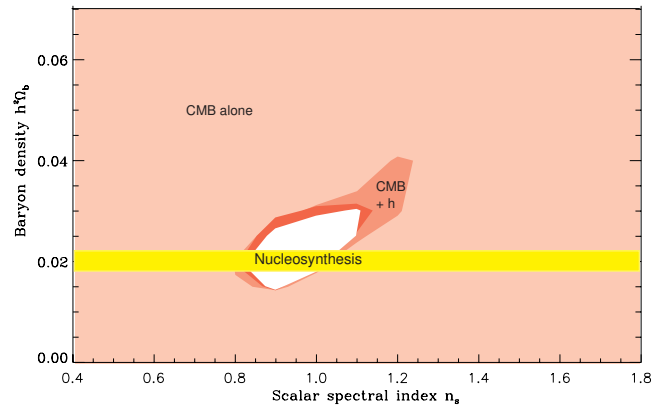}}
\bigskip
\caption{\label{nsobFig}\footnotesize%
Constraints in the $(\ns,\ob)$-plane.
The yellow/light gray band shows the BBN constraints from \protect\cite{Burles01}.
}
\end{figure}

\begin{figure}[tb] 
\centerline{\epsfxsize=8.5cm\epsffile{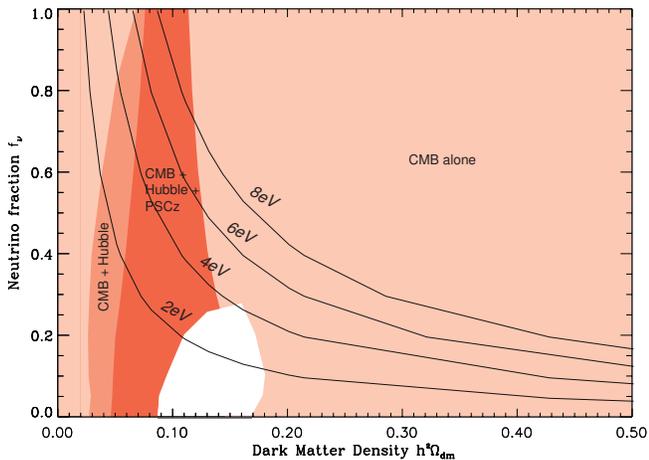}}
\bigskip
\caption{\label{odfnFig}\footnotesize%
Constraints in the $(\od,\fn)$-plane.
The four curves show contours of constant neutrino mass sum.
}
\end{figure}

\subsubsection{Constraints from CMB alone}

Figures \ref{odobFig}, \ref{nsobFig} and \ref{odfnFig} summarize some 
of the key 2-dimensional constraints on the matter budget. 
It is seen that the CMB data alone are now for the first time 
(compare, eg, \cite{concordance}) powerful enough to close
off bounded regions in these planes. 
The low second acoustic peak inferred from the first Antarctic BOOMERaNG
results \cite{deBernardis00} was still consistent with a purely
baryonic Universe, prompting speculation \cite{McGaugh00} that
an alternative theory of gravity might be able to eliminate the 
need to for dark matter altogether. Since the more accurate measurements
of the second peak height from the new BOOMERaNG, DASI and Maxima data
give a higher value, even the CMB alone now requires 
a non-zero amount of dark matter, at least at modest significance. 

The increased second peak height
is also seen to resolve a second hot discussion topic of the past 
year: the CMB lower limit on the baryon density has now dropped down
in beautiful agreement with the BBN prediction.

\begin{figure}[tb] 
\vskip-1.2cm
\centerline{\epsfxsize=9.0cm\epsffile{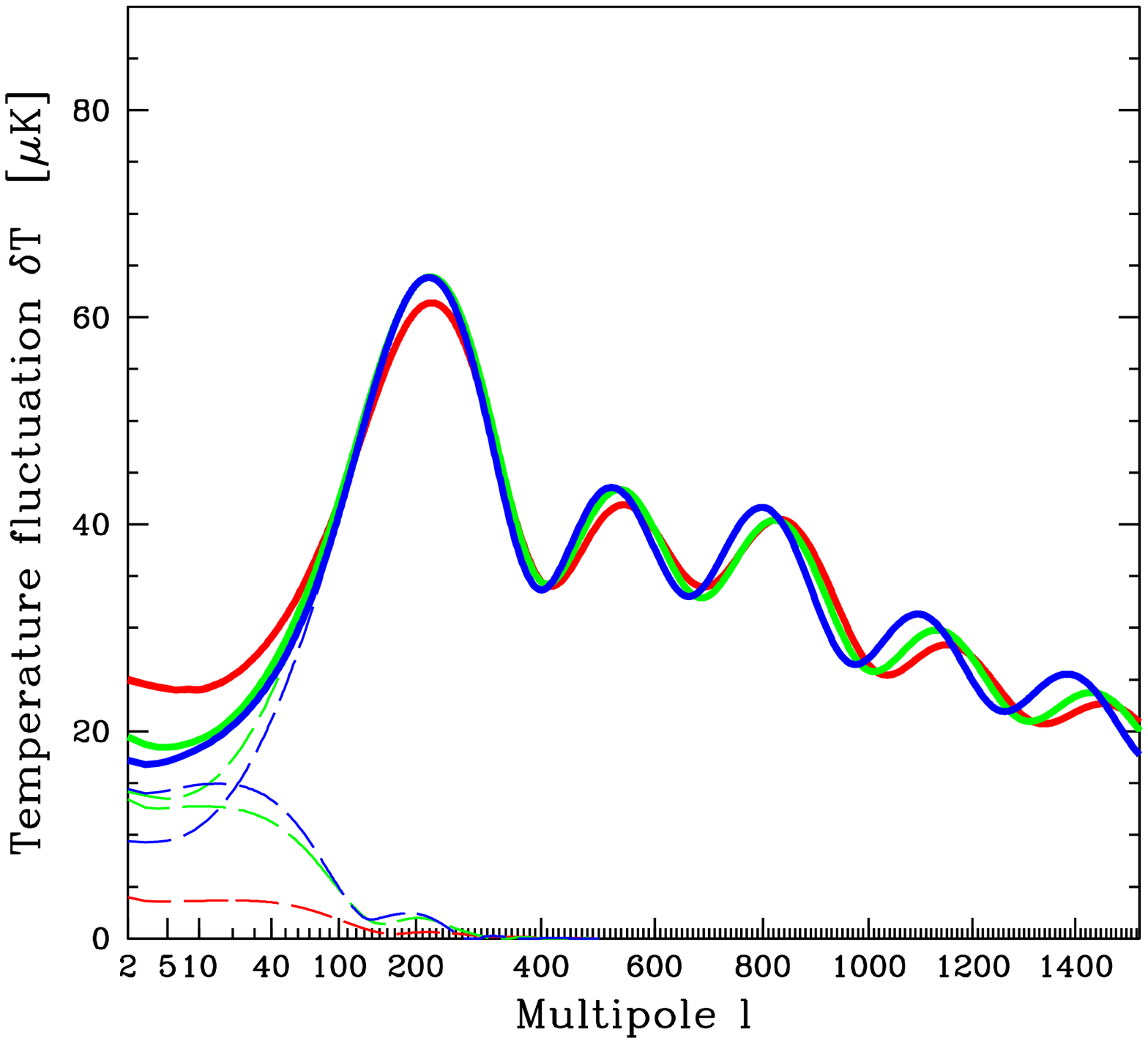}}
\vskip-0.8cm
\smallskip
\caption{\label{degenFig}\footnotesize%
The main remaining CMB degeneracy.
The three solid curves show the best fit models for CMB alone
subject to the constraint 
$\ob=0.02$ (red),
$\ob=0.03$ (green) and
$\ob=0.04$ (blue),
spanning the ``banana'' allowed by the CMB in \protect\fig{nsobFig}.
Dashed curves show the scalar and tensor contributions.
All models have $\tau=\fn=0$.
The parameters $\p\equiv(\Ok,\Ol,\od,\ob,\ns,\nt)$
are
$(-0.043,0.461,0.116,0.020,0.924,0.982)$,  
$(-0.134,0.799,0.050,0.030,1.200,0.884)$ and\hfill\\
$(-0.160,0.840,0.043,0.040,1.401,0.955)$, 
respectively.
}
\end{figure}

A third noteworthy result is that the allowed region in our
11-dimensional space is shaped like a long and skinny 
rather one-dimensional tube. This is seen clearly in the
projection onto the $(\ns,\ob)$ plane in \fig{nsobFig}.
The physics underlying this CMB degeneracy is illustrated in \fig{degenFig}.
Starting in the generally favored (white) region of 
\fig{nsobFig} and moving up to the right, 
the Universe becomes more and more closed ($\Ok$ decreases to negative values),
which would on its own shift the acoustic peaks too far to the left.
This is compensated by reducing the density of dark matter $\od$
and increasing the dark energy density $\Ol$ and 
the power spectrum tilt $\ns$ so that the peak location stays
essentially fixed. Through a rather spurious coincidence, 
the resulting changes in the various peak heights relative to the COBE
scale can be almost perfectly reversed by increasing the baryon density
and adding substantial amounts of gravity waves.
This is in essence the familiar $(\Ok,\Ol)$ angular distance degeneracy 
described in many parameter forecasting papers 
\cite{parameters,parameters2},
with the extra twist that changing $\ob$, $\od$, $\ns$ and $r$ 
as well helps minimize the change in the late integrated Sachs-Wolfe effect
on COBE scales.

\subsubsection{Breaking the CMB degeneracy}

Since this degeneracy involves so many parameters, prior constraints
on any one of them will help break it. In particular, the reason that
this degeneracy is not as prominent in the recent analyses by 
the BOOMERaNG \cite{Netterfield01}, DASI \cite{Pryke01} and 
Maxima \cite{Stompor01} teams 
is because they all assumed negligible gravity waves, $r=0$.

Since the Hubble parameter is given by 
\beq{hEq}
h = \sqrt{\od+\ob\over 1-\Ok-\Ol},
\eeq
it decreases sharply as one moves along the degeneracy track, dropping
as low as $0.3$ at the upper right endpoint in \fig{nsobFig}.
Imposing the prior $h=0.72\pm 0.08$ therefore shrinks
the allowed 11-dimensional region substantially, among other
things tightening the 
lower limit on dark matter in \fig{odobFig} 
and 
the upper limit on baryons in 
\fig{nsobFig}.

Because of its sensitivity to $\ns$ and $\od$ in particular, 
our galaxy clustering data (the PSCz power spectrum) is seen
to break this CMB degeneracy completely, shrinking the
allowed CMB ``bananas'' to almost round regions
in \fig{odobFig} and \fig{nsobFig}. 
The effect of adding PSCz is particularly dramatic in
\fig{odfnFig}, since CMB alone has almost no 
sensitivity to the neutrino fraction, whereas increasing 
$\fn$ provides a strong suppression of the galaxy power spectrum
on small scales.

The model best fitting the CMB+PSCz+$h$ constraints
has $\chi^2\approx 126$ for $105+32+1-11=127$ degrees of freedom.
The effective number of degrees of freedom might be a few larger than this,
since some of the 11 parameters had little effect, but
even if we ignore this, all fits are good in the sense of 
giving reduced $\chi^2$-values of order unity.

Adding our additional priors
$\ob=0.02$, $\tau=0$, $\Ok=0$ and $r=0$ cause little further
change, since they are all consistent with the favored
results and there are no major degeneracies left to break.

\begin{figure}[tb] 
\vskip-1.2cm
\centerline{\epsfxsize=9.0cm\epsffile{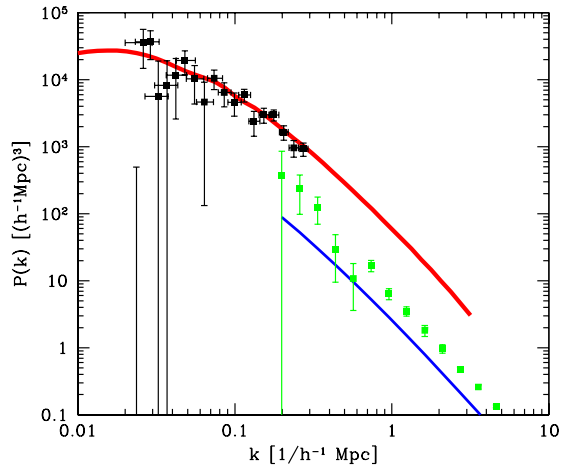}}
\vskip-0.8cm
\smallskip
\caption{\label{bestfitFig}\footnotesize%
The model best fitting the CMB, LSS and Ly$\alpha$F data
with $h$ and $\fn$ priors is shown together with the 
two data sets. The red and blue curves are the model predictions
for the LSS and Ly$\alpha$F data, respectively.
The blue curve is simply the red one shifted vertically and horizontally
to account for the fluctuation growth and 
Hubble parameter shift between $z\sim 2.72$ and the present epoch.
The overall calibration uncertainty in the 
Ly$\alpha$F data is seen to let the model prediction
lie below the data by a constant factor.
}
\end{figure}

Adding our Ly$\alpha$F data produced effects quite similar to those of
adding PSCz: breaking the CMB degeneracy and favoring 
a flat, roughly scale-invariant Universe. 
We have chosen to highlight the effects of the PSCz data rather those
of the Ly$\alpha$F data in the figures since we found 
its overall constraining power on parameters to be slightly 
stronger. For instance, the CMB+Ly$\alpha$F
constraints on tilt, curvature and baryon density are
\zz{0.91}{0.14}{0.10},
\zz{-0.03}{0.07}{0.22} and 
\zz{0.021}{0.008}{0.005}, respectively.
Overall, the PSCz and Ly$\alpha$F were found to be strikingly
consistent in pulling the CMB in the same direction, as was also 
found in \cite{Hannestad01}. 
Augmenting the CMB+PSCz data with the 13 Ly$\alpha$F points
increased the $\chi^2$ for the best fit model by only 10,
and the good agreement between the galaxy and Ly$\alpha$F
can also be seen visually, in \fig{bestfitFig}.

We conclude this subsection by summarizing what is obtained
by adding successively stronger assumptions as in Table 2.
\begin{enumerate}
\item CMB alone now gives constraints on most
parameters, but they are generally weak because of the 
above-mentioned degeneracy. 
\item Adding the PSCz galaxy clustering constraints 
breaks this degeneracy, resulting in tight 
constraints on everything except the dark energy density $\Ol$.
\item Adding the constraint $h=72\pm 8$ finally nails down $\Ol$.
It also sharpens the limits on $\Ok$ and $\fn$.
Adding this one constraint raises $\chi^2$ by as much as 4 (from 122
to 126), which reflects a slight tension between the $h$-prior 
and the CMB+PSCz data, which alone prefer the lower range 
$h=$\zz{0.57}{.31}{.30}.
\item Adding an $\fn=0$ prior boosts $\chi^2$ only by 0.1, 
further sharpening the $\Ok$, $\Ol$ and $\tau$ constraints slightly,
and none of our additional priors (including Ly$\alpha$F) have 
much of an effect. The only exception
is imposing $\ns=1$, which is slightly disfavored by the data and raises
$\chi^2$ by 4.
\end{enumerate}

\subsection{Inflationary parameters}

We now turn our attention to the parameters associated
with inflation. Since space remains perfectly consistent with
the inflationary flatness prediction despite the sharp 
error bar reduction caused by the new 
latest CMB data ($\Ok=0.00\pm 0.06$ at 95\% for CMB+PSCz+$h$),
it is interesting to quantify the constraints on the parameters $\ns$,
$\nt$ and $r$ to see how they compare with the predictions of 
various models, as has previously been done using 
earlier CMB data \cite{Kinney01,Hannestad01,Durrer00}.

\begin{figure}[tb] 
\centerline{\epsfxsize=8.5cm\epsffile{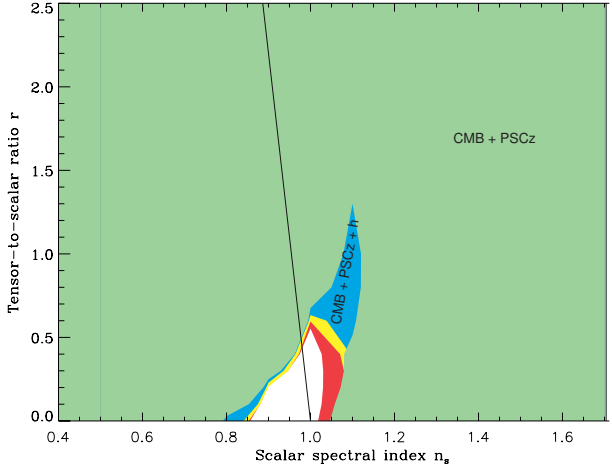}}
\bigskip
\caption{\label{nsrFig}\footnotesize%
Constraints in the $(\ns,r)$ plane.
The large green/light grey region shows the joint constraints from CMB and LSS. 
The blue/grey region shows the effect of adding the prior $h=0.72\pm0.08$.
The small yellow/light grey region shows the effect of adding $\fn=0$.
The red/dark grey region shows the effect of adding the BBN
prior $\ob=0.02$.
The straight line shows the prediction 
$r = (200/9)(1-\ns)$ 
from a power law class of inflation models.
}
\end{figure}

\begin{figure}[tb] 
\centerline{\epsfxsize=8.5cm\epsffile{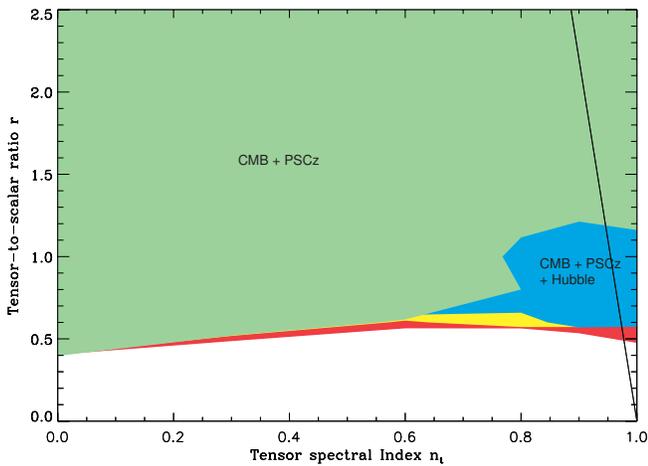}}
\bigskip
\caption{\label{ntrFig}\footnotesize%
Same as previous figure but for the 
$(\nt,r)$ plane.
The straight line shows the ``inflationary consistency relation''
$r = -(200/9)\nt$.
}
\end{figure}

As discussed in \cite{Martin00}, there has been a fair amount of
notational confusion in the literature surrounding the 
tensor-to-scalar ratio $r$.
There are two logical ways to define this ratio:
either in terms
of the fundamental parameters of the power spectrum (or, 
equivalently, of the inflationary model space),
or in terms of the observables, usually the CMB quadrupoles.
We adopt the former approach, and define 
\beq{RDefEq}
r\equiv {A_t\over A_s},
\eeq
where $A_s$ and $A_t$ are the scalar and tensor fluctuation amplitudes
as defined in \cite{Martin00}.
For inflation models where the slow-roll approximation is valid, 
this ratio is related to the tensor tilt $\nt$ by the so-called
inflationary consistency condition \cite{Martin00,Liddle92}
\beq{InfConsistencyEq}
r = -{200\over 9} \nt.
\eeq
A power law class of inflation models (see \cite{Kinney01} for a review)
make the additional prediction that
\beq{InfEq2}
r \equiv {200\over 9} (1-\ns).
\eeq
\ie, that $\ns = \nt + 1$.
(Although the quantity $\nt+1$ would be a more natural definition
for the tensor spectral index, we will stay true to the astronomical 
tradition of clinging to silly notation for historical reasons.) 

A common alternative definition of the tensor-to-scalar ratio is
the quadrupole ratio
\beq{rDefEq}
R \equiv {C^{\rm tensor}_2\over C^{\rm scalar}_2},
\eeq
In this case, the inflationary consistency condition is \cite{Martin00}
\beq{InfConsistencyEq3}
R \approx -{6.93} \nt
\eeq
for the special case where $\Ok=\Ol=0$.

Writing the relation between $R$ and $r$ as 
\beq{RrrelateEq}
r \approx a R,
\eeq
the proportionality constant $a$ is
typically between 2 and 5 --- it depends on the values of $\Ol$ and $\Ok$ 
via the late integrated Sachs-Wolfe effect. 

As shown in Table 2, \fig{1DnoFig} and \fig{1DhubbleFig},
the data prefers no tensor contribution at all, 
placing a 95\% upper limit 
$r<0.5$ for the CMB+PSCz+$h$ case. For comparison with prior work, 
this corresponds to a quadrupole ratio of 0.2 in the sense that
this is the quadrupole ratio for
the best fit model in our grid with this $r$-value (which is by
definition ruled out at exactly the 95\% level).
For comparison, previous multiparameter analyses 
incorporating gravity waves \cite{Kinney01,Hannestad01}
found $r<0.3$ using 
older CMB data and stronger priors.

Figures~\ref{nsrFig} and~\ref{ntrFig} show the joint constraints on r
with the scalar and tensor tilts, respectively, and allows comparison with
the predictions of equations\eqn{InfConsistencyEq} and\eqn{InfEq2}.
The constraints in the $(\ns,r)$ have becomes progressively sharper during
the past year \cite{Kinney01,Hannestad01}, and are now 
gradually becoming quite interesting.
In particular, 
\fig{nsrFig} shows that the preference for a slight red-tilt ($\ns\sim 0.9$)
and low $r$ in the data favors so-called ``small-field'' models \cite{Kinney01}.

\Fig{ntrFig} shows that the holy grail of testing the 
inflationary consistency relation\eqn{InfConsistencyEq}
is still a ways off. As expected, the constraints on the tensor tilt $\nt$ 
are stronger for high tensor normalizations $r$ and vanish completely
when $r$ does.

\section{Discussion}
\label{DiscussionSec}

In \sec{SystematicsSec}, we compared and combined the different CMB experiments,
finding that a consistent picture of the angular power spectrum emerged.
In \sec{ParameterSec}, we compared and combined a limited number of cosmological
data sets (power spectra from CMB, LSS, Ly$\alpha$F and various priors),
finding that a consistent set of cosmological parameters emerged that provided
a good fit to this data.

%

We conclude with some remarks on how these parameter measurements
match up with the many other cosmological observations that probe these parameters,
focusing on the cosmic matter budget.
In light of the checkered history of many cosmological parameters,
where tiny quoted error bars have repeatedly masked larger uncertainties
about underlying assumptions, such end-to-end consistency checks are crucial.
For instance, the Hubble parameter $h$ has dropped by a factor of eight since 
Hubble's original paper, 
the BBN baryon density $\ob$ has risen by 50\% in less than a decade,
and the favored value of the cosmological constant $\Ol$ has 
fluctuated wildly at the hands on both theorists and observers.

\subsection{Baryon density $\ob$}

The rise of the second peak in the new data has completely 
eliminated the tension between BBN and CMB, 
and they two are now in beautiful agreement that 
the baryon density $\ob\approx 0.02$.
This agreement was noted in the latest team papers as well
\cite{Netterfield01,Pryke01,Stompor01}.
That one method involving nuclear physics when the Universe was 
seconds old and another involving plasma physics more than 
100{,}000 years later give the same answers, despite involving
completely different systematics, is a landmark achievement
for cosmology. It greatly boosts the credibility of the
basic cosmological storyline since 
the Universe was a split second old.
This sudden agreement is all the more impressive given the 
lack thereof during the past year and the ado this generated.

Consensus has yet to be reached on the second decimal of the BBN 
predictions,
with the value $\ob=0.023$ from a recent VLT deuterium study \cite{Sandro01}
lying above the 95\% range of \cite{Burles01}.
As a reality check, our baryon value also agrees with 
slightly less accurate estimates of the 
abundance in the local Universe --- \eg, 
the range $0.007\simlt\Ob\simlt0.041$
inferred from a low-redshift inventory \cite{Fukugita98}
and the range
$0.015\simlt\ob\simlt 0.03$
at redshifts of a few from the Ly$\alpha$ forest
\cite{Rauch00,Hui01}, although the latter tend to be on the high side.
Similarly, the inferred baryon fraction $\ob/\od\sim 18\%$
agrees with that inferred from 
galaxy clusters \cite{Durrer00,Mohr00},
although this match weakly prefers lower $h$-values.

\subsection{Dark matter and dark energy density}

By now, there are a large number of independent methods for probing 
the dark matter density $\od$ or the (almost identical) total density, including 
studies of the cluster abundance at various redshifts
\cite{Mohr00,Diego01,Bahcall98,Eke98,Henry00},
mass-to-light ratio techniques \cite{Bahcall01,Carlberg99},
the baryon fraction $\Ob/\Om$ from cluster studies
\cite{Grego01}, 
cosmic velocity fields and
redshift space distortions.
Most of these probes do not measure $\od$ directly, but
$\Od=\od/h^2$ or some combination of this with the
fluctuation normalization parameter $\sigma_8$.
Although
each comes with potential systematic
errors of its own, 
and there is still some internal controversy in the velocity and 
cluster areas \cite{Liddle99}, 
they are all broadly consistent with 
the range $\od=0.09-0.18$ listed in Table 2.
Indeed, it is interesting that some of the last evidence 
supporting $\Om\sim 1$ came from velocity fields and 
redshift distortion studies, and that improved data now 
gives values as low as in Table 2 from both 
velocity fields \cite{Zehavi99,Nusser00} and
redshift space distortions \cite{pscz,Valentine00,Peacock01}.

With CMB strongly favoring an essentially flat Universe, 
the low matter density automatically implies a high dark energy 
density. As has been discussed in numerous papers
(\eg, \cite{Netterfield01,CosmicTriangle,concordance}),
there are now independent ways of reaching this conclusion,
making it robust to throwing away information on either
supernovae 1a \cite{Perlmutter98,Riess98} or large-scale structure.
The BOOMERaNG team \cite{Netterfield01} 
argue that even CMB and Hubble constant measurements
now favor $\Ol>0$ --- we have seen that this conclusion hinges on the the 
important additional assumption of negligible gravity waves.

\subsection{Neutrino density}

For the neutrino density $\on$, we are still a far cry from 
the grand goal of a precision cross-check between cosmological and laboratory
measurements, since the two have so far provided only upper and lower
limits, respectively. However, the two limits are steadily creeping closer.
Atmospheric neutrino oscillations \cite{Fukuda99} show that there is at least one
neutrino (presumably a linear combination of 
$\nu_\mu$ and $\nu_\tau$) whose mass exceeds a lower limit
somewhere in the range 0.04-0.08 eV \cite{Valle01}.
Since 
\beq{nuEq}
\on\equiv{1\over 94\eV}\sum_i m_i,
\eeq
where $m_i$ is the mass of the $i$th neutrino,
this corresponds to a 
lower limit $\on\simgt 0.0004-0.0008$, or 
$\fn\simgt 0.003-0.01$,
Our constraints (see \fig{odfnFig}) give $\sum m_i <4.2$eV,
further sharpening the 5.5 eV limit from a careful analysis of
previous cosmological data \cite{Croft99} (see also \cite{Durrer00}). 
The mass of the heaviest neutrino is thus in
the range $0.04-4.2$ eV. 

Moreover, the mass-splittigs indicated by both solar and atmospheric
neutrino data are much smaller than 4.2 eV, suggesting 
that all three mass eigenstates would need to be almost 
degenerate for neutrinos to weigh in near our upper limit.
This means that the upper limit on the (almost identical) masses 
of the three neutrino states would be $4.2/3 = 1.4$ eV.

Note that if, as current data suggests, the mixing between
the flavor eigenstates $\nu_e$, $\nu_\mu$ and $\nu_\tau$ is not small, 
it is inappropriate to identify the three mass-eigenstates 
$\nu_1$, $\nu_2$ and $\nu_3$ with these flavor eigenstates.
For instance, the heaviest eigenstate $\nu_3$ is likely to
be almost a $50-50$ mixture of $\nu_\tau$ and $\nu_\mu$. 
The correct way to phrase our upper limit is therefore 
as a 1.4 eV upper limit on the the mass of $\nu_3$.

Finally, a caveat about non-standard neutrinos is in order. 
To first order, our cosmological constraint probes  
only the {\it mass density} of neutrinos. Our conversion
of this into a limit on the mass sum assumes that the
neutrino number density is known and given by the standard model
freezeout calculation; 112 cm$^{-3}$. In more general scenarios with sterile
or otherwise non-standard neutrinos where the freezeout abundance 
is different, the robust conclusion to take away is 
simply an upper limit on the total light neutrino mass density
of $8.4\times 10^{-28}$kg/m$^3$.

How do our results compare with those of other recent analyses?
The analysis most similar to ours is that of the 
BOOMERaNG team \cite{Netterfield01}.
A detailed numerical comparison of their results with our Table 2 
is very encouraging. 
Despite major differences in both analysis technique
(priors, parameter space, marginalization method, \etc)
and data used (that analysis was limited to BOOMERAnG and DMR data),
both the central values and the error bars are very similar for
most parameters when taking into account that they and we quote 
$1\sigma$ and $2\sigma$ errors, respectively.
This indicates that
what is being measured is really
borne out loud and clear by the data in a way that is robust towards
data selection or analysis details. 
Perhaps we are inevitably approaching the
dreaded day when not only cosmology is consistent, but cosmologists are as well.


\bigskip
The authors wish to thank Ang\'elica de Oliveira-Costa,
Alexander Friedland, 
David Hogg, Andrew Jaffe, Josh Klein, John Kovac, Pat McDonald,
Jim Peebles, Clem Pryke, Adam Riess, Dominik Schwarz, 
George Smoot, 
David Weinberg,
and especially Andrew Hamilton for 
helpful comments and suggestions. 
Support for this work was provided by
NSF grant AST00-71213,
NASA grants NAG5-9194 and NAG5-11099,
the University of Pennsylvania Research Foundation,
the Zaccheus Daniel Foundation and two awards from
the David and Lucile Packard Foundation.

\appendix

\section{Linearized modeling of calibration and beam errors}

This appendix describes our modeling of the four terms of the
error matrix from \eq{Neq}, assuming that all relative
errors are small ($\ll 1$).

$\Cuc$ reflects the part of the errors which are uncorrelated
between the different experiments and is due to detector noise and 
sample variance. We approximate it by
\beq{uncorrCueq}
\Cuc_{ij}\equiv\delta_{ij}\sigma_i^2,
\eeq
where $\sigma_i$ is defined as the average of the upper and lower error
bars quoted for $d_i\equiv\dT^2$. 

The last three terms in \eq{Neq} reflect correlations between measurements due
to calibration and beam errors. As in \cite{10par,KnoxPage00}, $\Cic$ 
is the part specific to a single 
multi-band experiment and $\Csc$ is the part that is correlated with
other experiments that are calibrated off of the same (slightly uncertain) source.
TOCO, MSAM, CBI and BOOMERaNG all calibrate off of Jupiter.
To be conservative, we assume that the full 5\% 
calibration uncertainty from 
Jupiter's antenna temperature is shared by these experiment.
The true correlation should be lower, since the four experiments 
observed Jupiter at different frequencies.
The remaining multi-band experiments in \fig{cmbdataFig}
should not have any such inter-experiment correlations.
This contribution to the noise matrix is therefore
\beq{CscEq}
\Csc_{ij}\equiv (2s_{ij})^2 y_i y_j,
\eeq
where 
\beq{rEq}
s_{ij} = \cases{
5\%   &if $i$ and $j$ refer to a Jupiter-calibrated point,\cr
0     &otherwise.\cr
}
\eeq
The factor of 2 in \eq{CscEq} stems from the fact that the percentage 
error on $\dT_i^2$ is roughly twice that for $\dT_i$ as long as 
it is small. We simply use the observed values for $y_i$ in this expression.

Similarly, the remaining calibration term is 
\beq{uncorrCieq}
\Cic_{ij}\equiv (2r_{ij})^2 y_i y_j,
\eeq
where $r_{ij}=0$ if $i$ and $j$ refer to different experiments.
If band powers $i$ and $j$ are from the same experiment, then
$r_{ij}$ is the quoted calibration error with
the source contribution $s_{ij}$ subtracted off in quadrature.
We use 
10\% for QMASK, 
14\% for Python V,
8\% for Viper, 
8.7\% for Toco 97,
6.2\% for Toco 98,
10\% for QMASK,
6.4\% for BOOM97,
10\% for BOOM98,
4\% for Maxima and
4\% for DASI. 

A Gaussian beam of standard deviation $\theta$ suppresses power
by a factor well approximated by $e^{(\l\theta)^2}$.
Taylor expanding this expression shows that 
a small beam error $\Delta\theta$ causes 
the power to be mis-estimated by a percentage
$2\l^2\theta\Delta\theta$.
Defining  
\beq{bDefEq}
b_i\equiv 2 {\Delta\theta_i\over\theta_i} \left({\l\theta_i}\right)^2 y_i,
\eeq
we can therefore write the beam error as 
\beq{beamEq}
\Cbeam_{ij}= b_i b_j
\eeq
when $i$ and $j$ refer to the same experiment, zero otherwise.
We use this approximation for BOOM98, where $\theta=12.9'/c$,
$\delta\theta=1.4'/c$, and $c=\sqrt{8\ln 2}$ is the familiar FWHM conversion factor.
The other experiment reporting 
important beam uncertainties is Maxima, for which
we use the $b_i$-coefficients published in \cite{Lee01} in place
of the approximation of \eq{bDefEq}.




\ed
\begin{references}   %

\bibitem{Netterfield01}
\rfprep\nnn Netterfield C B;2001;astro-ph/0104460

\bibitem{Halverson01}
\rfprep\nnn Halverson N W;2001;astro-ph/0104489
     
\bibitem{Lee01}
\rfprep\nnn Lee A T {\etal};2001;astro-ph/0104459

\bibitem{PeeblesYu70}
\rf\nnnn Peebles P J E\dualand\nnn Yu J T;1970;ApJ;162;815 

\bibitem{Sunyaev70}
\rf\nn Sunyaev R\dualand\nn Zeldovich {Ya};1970;Astrophys. \& Space Sci.;7;3

\bibitem{Lange00}
\rf\nnn Lange A E {\etal};2001;Phys. Rev. D;63;042001

\bibitem{boompa}
\rf\nn Tegmark M\dualand\nn Zaldarriaga M;2000;Phys. Rev. Lett.;85;2240    

\bibitem{Bambi00}
\rf\nn Balbi A {\etal};2000;ApJL;545;L1

\bibitem{observables}
\rf\nn Hu W, \nn Fukugita M, 
\nn Zaldarriaga M\multiand\nn Tegmark M;2001;ApJ;549;669

\bibitem{Jaffe00}
\rf\nn Jaffe A {\etal};2000;Phys. Rev. Lett.;86;3475

\bibitem{Padmanabhan00}
\rf\nn Padmanabhan T\dualand\nnn Sethi S K;2001;ApJ;555;125

\bibitem{Lineweaver00}
\rfprep\nnn Lineweaver C H;2000;astro-ph/0011448



\bibitem{Kinney01}
\rf\nnn Kinney W H, 
\nn Melchiorri A\multiand\nn Riotto A;2001;Phys. Rev. D;63;23505

\bibitem{Hannestad01}
\rfprep\nn Hannestad S, \nnn Hansen S H, 
\nnn Villante F L\multiand\nnnn Hamilton A J S;2001;astro-ph/0103047

\bibitem{Hannestad01b}
\rf\nn Hannestad S;2001;Phys. Rev. D;64;083002

\bibitem{Griffiths01}
\rf\nnn Griffiths L M, \nn Melchiorri A\multiand\nn Silk J;2001;ApJ;553;L5

\bibitem{Phillips01}
\rf\nn Phillips J, \nnn Weinberg D H, \nnnn Croft R A C, \nn Hernquist L,
\nn Katz N, \nn Pettini M;2001;ApJ;560;15

\bibitem{BondConfProc}
\rfprep\nnn Bond J R {\etal};2000;astro-ph/0011378
     
\bibitem{CosmicTriangle}    
\rf\nn Bahcall N, \nnn Ostriker J P, \nn Perlmutter S\multiand\nnn Steinhardt P J;1999;Science;284;1481

\bibitem{Novosyadlyj00}
\rf\nn Novosyadlyj B, \nn Durrer R, \nn Gottl\"ober S,
\nnn Lukash V N\multiand\nn Apunevych S;2000;A\&A;356;418
		
\bibitem{Novosyadlyj00b}
\rfprep\nn Novosyadlyj B, \nn Durrer R, \nn Gottl\"ober S,
\nnn Lukash V N\multiand\nn Apunevych S;2000;astro-ph/0002522

\bibitem{Durrer00}
\rf\nn Durrer R\dualand\nn Novosyadlyj B;2001;MNRAS;324;560
        		
\bibitem{Bridle00}
\rf\nnn Bridle S L {\etal};2001;MNRAS;321;333


\bibitem{Turner01}
\rfprep\nnn Turner M S;2001;astro-ph/0106035

\bibitem{Holder01}
\rfprep\nn Holder G, \nn Haiman Z\multiand\nn Mohr J;2001;astro-ph/0105396



\bibitem{Padin01}
\rf\nn Padin S {\etal};2001;ApJL;549;L1

\bibitem{Efstathiou01}
\rfprep\nn Efstathiou G {\etal};2001;astro-ph/0109152

\bibitem{Lahav01}
\rfprep\nn Lahav O {\etal};2001;astro-ph/0112162

\bibitem{Pryke01}
\rfprep\nn Pryke C {\etal};2001;astro-ph/0104490
     
\bibitem{Stompor01}
\rfprep\nn Stompor R {\etal};2001;astro-ph/010506
     
     
\bibitem{10par}
\rf\nn Tegmark M\dualand\nn Zaldarriaga M;2000;ApJ;544;30 

\bibitem{qmask}
\rf\nn Xu Y, \nn Tegmark M, \nn{de Oliveira-Costa} A, 
\nn Devlin M, \nn Herbig T, \nnn Miller A D, 
\nnn Netterfield C B\multiand\nnn Page L A;2001;Phys. Rev. D;63;103002 


\bibitem{Netterfield97}
\rf\nnn Netterfield C B, \nnn Devlin M J, \nn Jarosik N, 
\nnn Page L A\multiand\nnn Wollack E J;1997;ApJ;474;47

\bibitem{qmap1}
\rf\nn Devlin M, \nn{de Oliveira-Costa} A, \nn Herbig T,
\nnn Miller A D, \nnn Netterfield C B, 
\nnn Page L A\multiand\nn Tegmark M;1998;ApJL;509;L77

\bibitem{qmap2}
\rf\nn Herbig T {\etal};1998;ApJL;509;L73

\bibitem{qmap3}
\rf\nn{de Oliveira-Costa} A, \nn Devlin M, \nn Herbig T,
\nnn Miller A D, \nnn Netterfield C B, 
\nnn Page L A\multiand\nn Tegmark M;1998;ApJL;509;L77

\bibitem{Gawiser00}
\rf\nn Gawiser E\dualand\nn Silk J;2000;Phys. Rept.;333-334;245
       
\bibitem{Bennett96}
\rf\nnn Bennett C L \etal;1996;ApJ;464;L1

\bibitem{qmaskpow}
\rfprep\nn Xu Y, \nn Tegmark M\multiand \nn{de Oliveira-Costa} A;2001;astro-ph/0104419 


\bibitem{ScottWhiteSilk95}
\rf\nn Scott D, \nn Silk J\multiand\nn White M;1995;Science;268;829

\bibitem{Pierpaoli00}
\rf\nn Pierpaoli E, 
\nn Scott D\multiand\nn White M;2000;Mod. Phys. Lett. A;15;1357
     
\bibitem{KnoxPage00}
\rf\nn Knox L\dualand\nn Page L;2000;Phys. Rev. Lett.;85;1366
     
\bibitem{BJK}
\rf\nnn Bond J R, \nnn Jaffe A H\multiand\nnn Knox L E;2000;ApJ;533;19

\bibitem{mapmaking}
\rf\nn Tegmark M;1997;ApJ;480;L87

\bibitem{Wright96}       
\rfprep\nnn Wright E L;1996;astro-ph/9612006

\bibitem{comparing}
\rf\nn Tegmark M;1999;ApJ;519;513


\bibitem{r}
\rf\nn Tegmark M\dualand\nnn Bromley B C;1999;ApJL;518;L69

\bibitem{EfstathiouNullbuster}
\rf\nn Seaborne M {\etal};1999;MNRAS;309;89

\bibitem{Press96}
\rfprep\nnn Press W H;1996;astro-ph/9604126:

\bibitem{concordance}
\rf\nn Tegmark M, 
\nn Zaldarriaga M\multiand\nnnn Hamilton A J S;2001;Phys. Rev. D;63;43007



\bibitem{Taruya01}
\rf\nn Taruya A, \nn Magara H, \nnn Jing Yi P\multiand\nn Suto Y;2001;PASJ;53;155

\bibitem{Saunders00}
\rf\nn Saunders W {\etal};2000;MNRAS;317;55
 
\bibitem{pscz}    
\rf\nnnn Hamilton A J S, 
\nn Tegmark M\multiand\nn Padmanabhan N;2000;MNRAS;317;L23

\bibitem{Croft00}
\rfprep\nnn Croft A C {\etal};2000;astro-ph/0012324
     
\bibitem{McDonald00}
\rf\nn McDonald P, \nn {Miralda-Escud\'e} J, \nn Rauch M,
\nnnn Sargent W L W, \nnn Barlow T A, 
\nn Cen R\multiand\nnn Ostriker J P;2000;ApJ;543;1
   
\bibitem{Hamilton00}
\rf\nnnn Hamilton A J S;2001;MNRAS;322;419

\bibitem{Burles01}
\rf\nn Burles S, \nnn Nollett K M\multiand \nnn Turner M S;2001;ApJ;552;L1

\bibitem{Freedman00}
\rf\nnn Freedman W L {\etal};2001;ApJ;553;47


\bibitem{Fukuda99}
\rf\nn Fukuda Y {\etal};1999;Phys. Rev. Lett.;82;1810

\bibitem{Valle01}
\rfprep\nnnn Valle J W F;2001;astro-ph/0104085

\bibitem{deBernardis00}
\rf\nn {de Bernardis} P {\etal};2000;Nature;404;955




\bibitem{McGaugh00}
\rf\nnn McGaugh S S;2000;ApJL;541;L33

\bibitem{parameters}
\rf\nnn Bond J R, \nn Efstathiou G\multiand \nn Tegmark M;1997;MNRAS;291;L33

\bibitem{parameters2}
\rf\nnn Eisenstein D J, \nn Hu W\multiand \nn Tegmark M;1999;ApJ;518;2


\bibitem{Martin00}
\rf\nn Martin J\dualand\nn Schwarz D;2000;Phys.Rev. D;62;103520

\bibitem{Liddle92}
\rf\nnn Liddle A R\dualand\nnn Lyth D H;1992;Phys. Lett. B;291;391

\bibitem{Sandro01}
\rf\nn {D'Odorico} S, \nn {Dessauges-Zavadsky} M\multiand\nn Molaro P;2001;A\&A;368;L21

\bibitem{Fukugita98}
\rf\nn Fukugita M, \nnn Hogan C J\multiand\nnnn Peebles P J E;1998;ApJ;503;518

\bibitem{Rauch00}
\rf\nn Rauch M {\etal};1997;ApJ;489;7

\bibitem{Hui01}
\rf\nn Hui L, \nn Haiman Z, 
\nn Zaldarriaga M\multiand\nn Alexander T;2002;ApJ;564;525
    



\bibitem{Mohr00}
\rfprep\nnn Mohr J J, \nn Haiman Z\multiand\nnn Holder G P;2000;astro-ph/0004244 

\bibitem{Diego01}
\rfprep\nnn Diego J M {\etal};2001;astro-ph/0104217

\bibitem{Bahcall98}
\rf\nnn Bahcall N A\dualand\nn Fan X;1998;ApJ;504;1
    
\bibitem{Eke98}
\rf\nnn Eke V R, \nn Cole S, \nnn Frenk C S\multiand\nnn Henry J P;1998;MNRAS;298;1145

\bibitem{Henry00}
\rf\nnn Henry J P;2000;ApJ;534;565

\bibitem{Bahcall01}
\rf\nnn Bahcall N A {\etal};2000;ApJ;541;1
     
\bibitem{Carlberg99}     
\rf\nnn Carlberg R G {\etal};1999;ApJ;516;552

     
\bibitem{Grego01}
\rf\nn Grego L {\etal};2001;ApJ;552;2

        
\bibitem{Liddle99}
\rf\nnn Liddle A R\dualand\nnnn Viana P T P;1999;MNRAS;303;535

\bibitem{Zehavi99}
\rf\nn Zehavi I\dualand\nn Dekel A;1999;Nature;401;252

\bibitem{Nusser00}
\rf\nn Nusser A, \nnn {da Costa} L N, \nn Branchini E, \nn Bernardi M, \nnn Alonso M V,
\nn Wegner Gary, \nnnn Willmer C N A, \nnn Pellegrini P S;2001;MNRAS;320;L21 

\bibitem{Valentine00}
\rf\nn Valentine H, \nn Saunders W\multiand\nn Taylor A;2000;MNRAS;319;L13

\bibitem{Peacock01}
\rf\nnn Peacock J A {\etal};2001;Nature;410;169


\bibitem{Perlmutter98}
\rf\nn Perlmutter S {\etal};1998;Nature;391;51

\bibitem{Riess98}
\rf\nnn Riess A G {\etal};1998;Astron. J.;116;1009

\bibitem{Croft99}
\rf\nnnn Croft R A C, \nn Hu W\multiand\nn {Dav\'e} R;1999;Phys. Rev. Lett.;83;1092










\end{references}
